Transparent Tiles of Silica Aerogels for High-Energy Physics

Makoto Tabata[a*,1]

[a] Department of Physics, Graduate School of Science, Chiba University, 1-33 Yayoicho, Inage-ku, Chiba 263-8522, Japan

[1] E-mail: makoto@hepburn.s.chiba-u.ac.jp.

Abstract

Silica aerogels are important to be used as photon radiators in Cherenkov counters for high-energy-physics experiments because of their optical transparency and intermediate refractive indices between those of gases and liquids or solids. Cherenkov counters that employ silica aerogels as radiators and photodetectors are often used to identify subatomic charged particles (e.g., electrons, protons, and pions) with momenta on the order of sub-GeV/$c$ to GeV/$c$; they are also used to measure particle velocities in accelerator-based particle- and nuclear-physics experiments and in space- and balloon-borne experiments in the field of cosmic-ray physics. Recent studies have demonstrated that it is important for the design of Cherenkov counters that the transparent silica-aerogel tiles comprise solid material with recently improved transparency and a refractive index that can be controlled between 1.003 and 1.26 by varying the bulk density in the range of 0.01–1.0 g/cm$^3$. Additionally, a technique for fabricating large-area silica-aerogel tiles without cracking has been developed. In this chapter, we describe advances in the technologies for producing silica aerogels with high optical performances to be used in scientific instruments. We further discuss the principles underlying the operation of detectors based on the Cherenkov effect. We also review applications of silica aerogels in specific high-energy-physics experiments.

1. Introduction

    Silica aerogels, a subset of the aerogel family, were first invented by S. S. Kistler in 1931 [1]. Since then, their attractive physical and chemical properties have led to their utilization in various applications. One of the successful applications is in Cherenkov light radiators, which are used in high-energy-physics research including accelerator-based particle- and nuclear-physics experiments and observational cosmic-ray (astroparticle) physics.

    In these fields of research, Cherenkov counters have been often used to detect, count, identify, and measure the velocity of high-energy subatomic charged particles with energies ranging from mega-electron volts (MeV) to tera-electron volts (TeV) (i.e., close to the speed of light). Cherenkov counters basically comprise optical radiators, photodetectors, and readout electronics. The optical radiators emit photons due to the passing of charged particles based on the Cherenkov effect. This Cherenkov radiation relies on the velocity of the particle and refractive index, $n$, of the radiator (as further detailed in Section 2.1). Thus, the Cherenkov radiators must be selected according to the physics objectives of each experiment. Pressurized gases (e.g., nitrogen ($N_2$), carbon dioxide ($CO_2$), and perfluorobutane ($C_4F_{10}$; $n \approx 1.00137$)), liquids (e.g., water; $n = 1.333$), and solids (e.g., lead glass) are usually used as radiators.

However, there is a large gap between the refractive indices of gases and liquids, leaving a needed for materials with refractive indices within this gap.

Silica aerogels are some of most the useful and promising materials as the Cherenkov radiators due to their transparency and adjustable refractive indices that can fill this gap between those of gases and liquids. This application was first reported by M. Cantin et al. in 1974 [2]. After that, researchers have been working to develop high-performance silica-aerogel tiles (Figure 1). Thus, this application of silica aerogels as Cherenkov radiators is considered traditional by the high-energy-physics community. In this chapter, we overview the major progress in the development of silica-aerogel radiators since the publication of the first edition of this handbook [4]. Herein, silica aerogels are referred to simply as "aerogels."

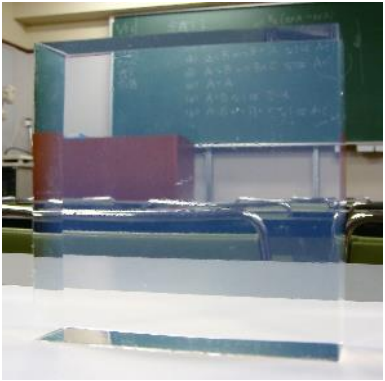

Figure 1. Highly transparent silica-aerogel tile produced at Chiba University with approximate dimensions of 9 × 9 × 2 cm and a refractive index of 1.065 at a density of 0.22 g/cm$^3$ (adapted from [3] ©The Chemical Society of Japan).

2. Aerogel Cherenkov Counters

2.1. Physics of the Cherenkov Effect

The Cherenkov effect was experimentally discovered by P. A. Čerenkov in 1934 [5,6] and theoretically explained by I. M. Frank and I. E. Tamm in 1937 [7,8]. The Nobel Prize in Physics 1958 was awarded jointly to the three scientists "for the discovery and the interpretation of the Cherenkov effect" [9].

The Cherenkov effect is a phenomenon involving rapid light emission and is briefly explained as follows. In a medium with a refractive index of $n$ (hereafter referred to simply as the "index"), the speed of light is reduced to $c/n$, where the constant $c$ is the speed of light in a vacuum. A sufficiently high-energy subatomic particle can travel with a velocity of $v$ exceeding that of light in the medium:

$v > c/n,$ (1)

which is consistent with Einstein's theory of special relativity. If the particle has electric charges and the medium is dielectric, a type of shock wave of electromagnetic radiation, called Cherenkov radiation, is emitted at an angle (the Cherenkov angle) to the direction of the particle's motion (Figure 2). Thus, a threshold velocity,

$$v_t = c/n, (2)$$

exists as a condition of the Cherenkov radiation. Letting

$$\beta \equiv v/c, (3)$$

the Cherenkov angle can be expressed as

$$\cos\theta_C = 1/n\beta. (4)$$

The Cherenkov radiation forms a ring (or disk) pattern on the plane perpendicular to the particle's direction of travel. From Equation (2), it can be said that the higher the velocity of the particle and the index of the Cherenkov medium, the larger the Cherenkov angle will be and, therefore, the larger this ring will be.

Consider the Cherenkov effect from the perspective of quantum mechanics, the Cherenkov light radiation can be thought of as Cherenkov photon emission in the $\theta_C$ direction. The number of emitted Cherenkov photons, $N$, at a wavelength between $\lambda$ and $\lambda + d\lambda$ is obtained from the following equation:

$$\frac{d^2N}{dxd\lambda} = \frac{2\pi\alpha z^2}{\lambda^2}\left(1 - \frac{1}{n^2\beta^2}\right), (5)$$

where $x$ is the pass length of the particle in the radiators, $\alpha$ is the fine-structure constant, and $z$ is the electric charge of the particle. Note that the shorter the wavelength is and the larger the thickness and the index of the radiator are, the more photons are emitted. Thus, the Cherenkov effect is a phenomenon of very faint light emission as the number of emitted photons is approximately two orders of magnitude smaller than that of scintillating light emission.

Cherenkov counters are one type of detector for subatomic charged particles (or ionizing radiation) and they employ the Cherenkov effect. The subatomic particles that can be detected using Cherenkov counters include leptons (such as electrons and muons), mesons (such as charged pions and kaons), various nuclei (including protons and deuterons), and their antiparticles. Cherenkov counters comprise, at a minimum, Cherenkov radiators, photodetectors, and signal readout electronics and sometimes contain mirrors to collect light. As described in the Introduction, transparent gases and condensed materials (i.e., liquids and solids), as well as aerogels, can be used as the Cherenkov radiator depending on the velocity range expected in a particular high-energy experiment. Two main types of Cherenkov counters have been employed in high-energy physics: The first is a threshold-type Cherenkov counter, which separates two kinds of particles or limits the velocity range of the particles by determining whether each particle fulfills the condition for Cherenkov radiation. The other is a ring-imaging Cherenkov (RICH) counter, which determines each particle's velocity by measuring its Cherenkov angle, which can be used to identify the types of particles. These are further described in the following sections.

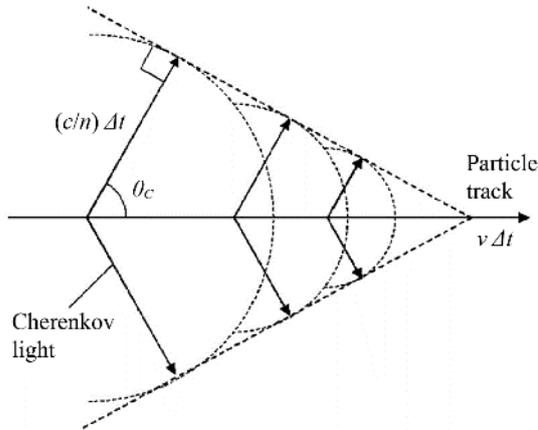

Figure 2. Schematic of Cherenkov radiation, where $\Delta t$ is the time for which the particle passes through the medium (adapted from [4] ©Springer).

2.2. Threshold-Type Cherenkov Counters

Threshold-type Cherenkov counters apply the threshold condition of the Cherenkov effect to identify the particle types (which is commonly called particle identification or PID) or to conduct velocity measurements. Hereafter, we assume that the Cherenkov radiators are aerogels and the corresponding threshold-type Cherenkov counters are generally called "aerogel Cherenkov (AC) counters." In high-energy physics experiments, the types and momenta of particles to be observed are sometimes known in advance. The types of an observed particle influences its target physics phenomena and its momentum can be measured by tracking its trajectory using a spectrometer operated in a magnetic field. Based on the particle's mass, $m$, the relativistic momentum, $p$, is defined as

$$p = \frac{mv}{\sqrt{1-\frac{v^2}{c^2}}}. \quad (6)$$

Choosing $n$ as the aerogel index, the threshold momentum, $p_t$, for fulfilling the Cherenkov condition can be expressed based on Equations (2) and (6) as follows:

$$p_t = \frac{m}{\sqrt{n^2-1}}, \quad (7)$$

where the natural system of units is used (i.e., $c = 1$). When $p$ exceeds $p_t$, Cherenkov light is emitted from the aerogel (which has a given refractive index, $n$) due to the passage of particles (AC = yes). Otherwise, Cherenkov light is not emitted (AC = no) and, thus, the photodetectors do not output any signal.

If two particle types have different masses and but identical momenta (or momenta limited within a certain range), an aerogel can be selected with an index that causes only the lighter particles trigger Cherenkov radiation. Consequently, PID can be achieved by measuring the photodetector output (i.e., AC = yes or no). This is the operational principle of threshold-type AC counters. As a representative example, Figure 3 shows the threshold momenta as functions of the index for three of the most common particles to be identified—pions, kaons, and protons, which have masses of 139.6, 493.7, and 938.3 MeV/$c^2$, respectively [10]. For a given momentum, note that the threshold index, $n_t$, can be derived from Equation (7). Assuming the conditions in which pions and kaons with $p = 2$ GeV/$c$ pass through an AC counter, the threshold indices are $n_t = 1.0024$ and $1.0300$ for pions and kaons, respectively. Thus, the condition for AC = yes for pions but AC = no for kaons can be created by setting the aerogel index to a value between 1.0024 and 1.0300. Practically, physicists choose the large index possible (in this case, approximately 1.03) to obtain high photon yields based on Equation (5).

AC counters should be designed to maximize the photoelectron yield in the photodetectors. The AC = yes/no determination depends largely on the (average) number of detected photoelectrons. Photodetectors must be able to detect a few as a single photoelectron; thus, photomultiplier tubes (PMTs) are used in most cases. The counter performance strongly relies on its design, including the Cherenkov light emittance (i.e., the optical quality and thickness of the aerogel) and light-collection efficiency (i.e., the counter size and geometry, the quantum efficiency, size, and total number of PMTs used, and the reflectance values of the mirror or diffuse reflectors). An example of the structure of an AC counter is shown in Figure 4. In this system, Cherenkov photons travel inside the AC counter box as they are scattered by the aerogels and reflected by the reflectors on the counter walls until they are absorbed by an aerogel or a reflector or detected by a photodetector. Consequently, the average number of photoelectrons detected is typically about 10 photoelectrons.

AC counters are useful as on-line event triggers or as veto devices at the hardware level. Recent advances in accelerator technology have enabled physicists to conduct high-intensity physics experiments, which require control over the particle-interaction phenomena at high event rates. To facilitate data acquisition, AC counters can be designed so that only interesting events triggered AC = yes, or background events are rejected as AC = yes (veto). Of course, the raw data can alternatively be analyzed off-line to judge AC = yes or no. Further, by using two or more AC counters with different indices, three or more different particle types can be distinguished. Similarly, velocity measurements in certain ranges for a specific types of particles can also be achieved.

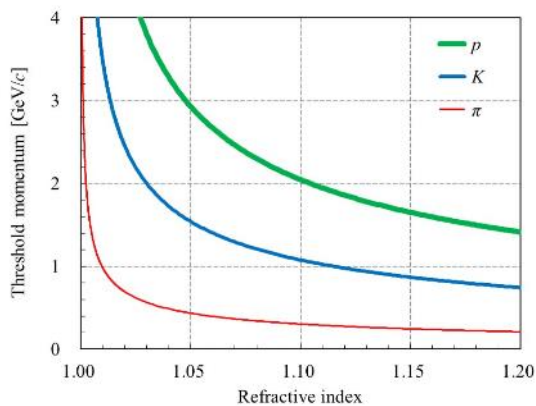

Figure 3. Threshold momenta as functions of the index for pions ($\pi$), kaons ($K$), and protons ($p$).

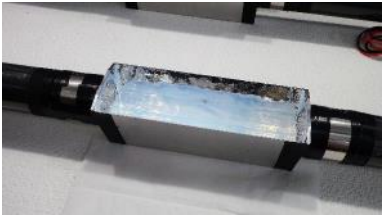

Figure 4. Example of an AC counter (the top cover is not shown): a light-shielded aluminum counter box is lined with aluminized mylar sheets (i.e., mirror reflectors) and filled with aerogel blocks; PMTs are attached to each of the longitudinal sides of the counter box.

2.3. Ring-Imaging Cherenkov Counters

RICH counters measure the particle velocity by imaging the Cherenkov light rings emitted from the radiators using position-sensitive photodetectors (see Figure 5). Assuming the use of aerogels as the radiators, these counters are referred to as "aerogel RICH counters." Using Equations (3) and (4), the particle velocity can be expressed as

$$v = \frac{c}{n\cos\theta_C}. \quad (8)$$

For a given aerogel index, the particle velocity can be measured based on the Cherenkov angle, which can be calculated by reconstructing the Cherenkov rings and determining their radii using a computer. Photodetectors are placed in downstream from the aerogels as shown in Figure 5. They must be pixelated (e.g., with a pixel size of 5 mm) and be able to count single photons because typical RICH counters observe Cherenkov rings of approximately 10 photoelectrons (see Figure 6). Using Equations (6) and (8), the particle mass can be expressed as follows:

$$m = \frac{p}{c}\sqrt{n^2 \cos^2\theta_C - 1}. \quad (9)$$

Thus, the particle mass can be determined by measuring the particle momentum and the Cherenkov angle (i.e., the particle velocity) using tracking devices and aerogel RICH counters, respectively, for a given aerogel index. This process based on Cherenkov angle measurements represents the essential principle of PID for particles with different masses with aerogel RICH counters.

Focusing RICH counters generate a Cherenkov ring with a finite radius on the photodetection plane (i.e., the surface of the photodetectors) by limiting the aerogel thickness (i.e., incorporating an expansion distance between the aerogels and the photodetectors) (see Section 33.5 in *Review of Particle Physics* [13]). Maintaining a long distance between the aerogels and photodetectors is helpful to clearly separate ring images from the photons associated with different particles as it encourages the formation of a ring image rather than a disk image. Therefore, physicists often place mirrors downstream from the aerogels to extend the distance and focus the Cherenkov photons on photodetectors; such counters are called mirror-focusing RICH counters. Alternatively, the photodetectors can be positioned behind a short expansion distance as shown in Figure 5 for cases such as when the counter installation space is limited; such counters are called proximity-focusing RICH counters.

The performance of a focusing RICH counter can be roughly estimated from the number of detected photoelectrons, $N_{p.e.}$, and the Cherenkov angle resolution for a single photon, $\sigma_\theta$, as

$$\frac{\Delta\theta_C \sqrt{N_{\text{p.e.}}}}{\sigma_\theta}, \quad (10)$$

where $\Delta\theta_C$ is the difference between the Cherenkov angles of two charged particles to be distinguished. The value $\sigma_\theta/\sqrt{N_{\text{p.e.}}}$ is referred to as the Cherenkov angle resolution per (charged-particle) track or per ring. Unlike with threshold-type Cherenkov counters, Cherenkov light must be emitted for both the particles in order for the particles to be identified using a RICH counter. For example, researchers can choose an aerogel with $n = 1.05$ to distinguish pions and kaons at $p > \sim 2$ GeV/$c$ (refer back to Figure 3). In this case, $\Delta\theta_C$ is approximately 23 mrad at $p = 4$ GeV/$c$. Assuming that $N_{\text{p.e.}} = 10$, a pion/kaon separation capability of $4\sigma$ is possible when $\sigma_\theta < 18.2$ mrad, which is conceivable considering the recent performances of aerogels and photodetectors.

To maximize $N_{\text{p.e.}}$, there is a needed for thicker and higher-index aerogels based on Equation (5). However, an optimal aerogel thickness exists that optimizes $\sigma_\theta$ by reducing the uncertainty of the Cherenkov photon emission points within the aerogel volume. Aerogel transparency is of particular importance because only photons that remain unscattered inside the aerogels retain the Cherenkov angle that is characteristic of the particle's properties. The value $\sigma_\theta$ can be evaluated by accumulating events from known particles (e.g., in a test-beam experiment), as shown in Figure 7. In the practical applications of the counter, PID can be carried out using, for example, maximum likelihood analysis [15,16].

In general, RICH counters have an advantage of being compatible with a wide momentum range compared to threshold-type counters apart from the yes/no decision. Moreover, RICH counters can resolve the problem of misidentification that occurs in threshold-type counters due to knock-on electrons (i.e., when lower-velocity particles that should trigger AC = no, occasionally collide with electrons inside the aerogels or counter box materials and, since such an electron can exceed the Cherenkov threshold, the counter is triggered as AC = yes). This issue is mitigated in RICH counters because knock-on electrons have different trajectories and momenta than the original particles.

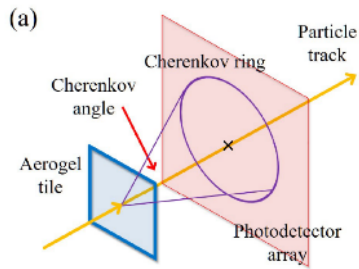

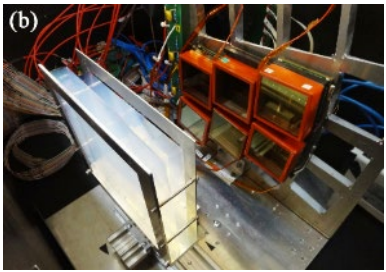

Figure 5. (a) Conceptual diagram of RICH counters (adapted from [3] ©The Chemical Society of Japan) (b) Prototype of a RICH counter used in a test-beam experiment on a partial model of the RICH detector used for the Belle II experiment, which is described in Section 4.2 (reproduced from [11] ©Elsevier B.V.).

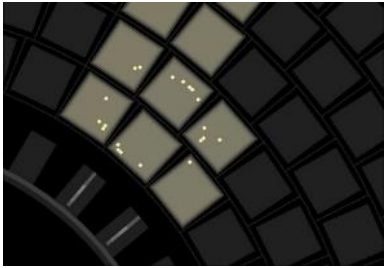

Figure 6. Cherenkov ring detected by a RICH counter as viewed on an event display reconstructed using a computer during a cosmic-ray test of the partial RICH detector module for the Belle II experiment, which is described in Section 4.2; the highlighted squares show the pixelated photodetectors used in the experiment and each Cherenkov photon (and background) hit derived from a cosmic-ray particle (probably a muon) is indicated by a yellow dot; aerogel tiles (not shown) were placed at the zenith position of the photodetectors (adapted from [12] ©Elsevier B.V.).

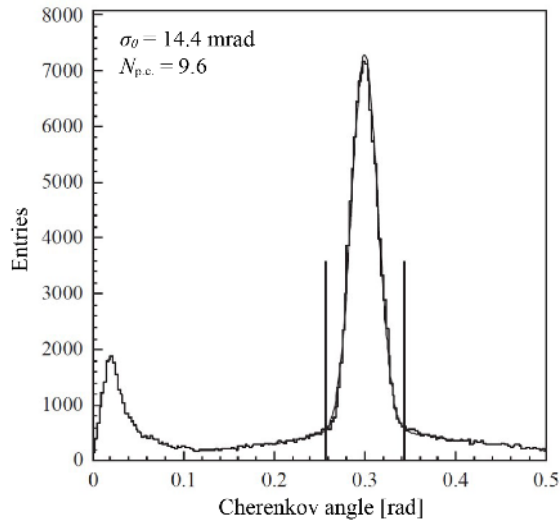

Figure 7. Example of the accumulated Cherenkov angle distribution of the hits of single Cherenkov photons (adapted from [14] ©Elsevier B.V.).

## 3. Transparent Silica-Aerogel Tiles

### 3.1. Production Methods and Activities in Russia and Japan

While many supply companies and research laboratories all over the world are studying aerogels, the centers of excellence for aerogel studies in the field of high-energy physics are limited to two groups at present: the Novosibirsk group in Russia and the Japanese group. Collaborating researchers from the Boreskov Institute of Catalysis and the Budker Institute of Nuclear Physics in Novosibirsk, Russia have been developing Cherenkov counters and the aerogels used the radiators since 1986 [17]. Similarly, the development of aerogels for use as Cherenkov radiators in Japan began around 1980 [18].

3.1.1. Russian Group

The aerogel fabrication procedures used by the Novosibirsk group can be divided into four stages: alcogel synthesis, aging, supercritical drying, and baking [17,19]. The alcogel preparation involves a variant of the two-step sol–gel method. First, an oligomer is synthesized via an acid-catalyzed reaction of tetraethoxysilane [TEOS, $Si(OC_2H_5)_4$] with water. Then, a solution of the oligomer and an excess of water are used for the base-catalyzed synthesis of alcogels. The alcogel aging stage is carried out in an alcohol bath, in which the alcohol is periodically replaced to wash out the water and the catalyst from the alcogel. To remove the organic solvent, high-temperature supercritical drying is performed using an autoclave at a temperature of 280°C and a pressure of 120 bar (12 MPa). The Novosibirsk group has two autoclaves: one with a 4 L volume that is used for test experiments and the other with a 100 L volume that is used for mass production. Finally, the aerogels are baked in air at 600°C to remove remaining organic impurities and water. The Novosibirsk aerogels are generally hydrophilic but, of course, they sometimes fabricate hydrophobic aerogels, as well [20]. The group and others have investigated the degradation of the optical properties of hydrophilic aerogels due to moisture absorption during high-energy-physics experiments as well as recovery procedures to counteract this process [21,22].

3.1.2. Japanese Group

In the early years of their research, Japanese physicists learned the aerogel fabrication technique from the Saclay group in France [2,18]. Several university physicists came together to develop aerogel Cherenkov counters at the High-Energy Accelerator Research Organization (KEK; formerly known as High-Energy-Physics Laboratory), in Tsukuba, Japan, which has since served as a hub for aerogel development in Japan. Physicists from the KEK group collaborated with experts in materials science from the Panasonic Corporation (formerly known as Matsushita Electric Works, Ltd. prior to being renamed as Panasonic Electric Works Co., Ltd.) in Osaka, Japan. Consequently, hydrophobic aerogels specifically designed for Cherenkov radiators were developed in the early 1990s [23,24]. Thereafter, the aerogels fabricated by the Japanese group are largely hydrophobic, as shown in Figure 8.

Based on the experimental production results, Panasonic Electric Works commercialized hydrophobic aerogel tiles with $n$ = 1.015, 1.03, and 1.05 (under the product names of SP-15, SP-30, and SP-50, respectively) around 1997 [4]. The standard dimensions of the Panasonic aerogel tiles were approximately 11 × 11 × 1 cm; however, tiles with different dimensions and/or indices are available on demand (e.g., SP-25 with $n \approx 1.025$ [26]). Since then, the Panasonic tiles have been considered the global standard for hydrophobic aerogels in Cherenkov radiators. Regrettably, Panasonic Electric Works discontinued their aerogel manufacturing around 2011 and the production technology was transferred to the Japan Fine Ceramics Center (JFCC) in Nagoya [27], which has taken over as the main supplier of commercial hydrophobic aerogels.

The aerogel fabrication procedure used by the KEK–Panasonic/Matsushita group can be divided into five stages: wet-gel synthesis, aging, hydrophobic treatment, solvent exchange, and supercritical drying [24,25]. The KEK–Panasonic group utilizes a variant of the two-step sol–gel method. The first step in the process used by the Russian group is omitted by using a commercially-available silica oligomer precursor, polymethoxysiloxane (PMS, also known as methyl silicate 51, $CH_3O[Si(OCH_3)_2O]_nCH_3$). To synthesize wet gels via the sol–gel process, PMS is mixed with water in either methanol or ethanol as a diluent/solvent and a base-catalyzed reaction is initiated with aqueous ammonia. In the early 2000s, *N*,*N*-dimethylformamide (DMF) was introduced as the new diluent/solvent for the wet-gel synthesis step [28]. After aging, the wet gels are immersed in alcohol and subsequently rendered hydrophobic by a surface modification using hexamethyldisilazane (HMDS, $[(CH_3)_3Si]_2NH$) [23] (see Figure 9):

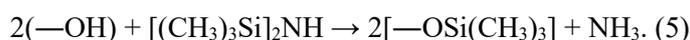

$$2(—OH) + [(CH_3)_3Si]_2NH \rightarrow 2[—OSi(CH_3)_3] + NH_3. \quad (5)$$

The alcohol solvent is repeatedly replaced to remove the impurities from the wet gels. Finally, the wet gels are dried by supercritical carbon-dioxide extraction. The aerogels are not baked to preserve the hydrophobic groups.

The KEK–JFCC/Panasonic group and the Chiba group employ three supercritical extraction systems. The oldest one, equipped with a 31 L-volume autoclave, seems to have been placed at KEK around 1980 for use in supercritical ethanol drying but may have been built in another laboratory before then. The second system is a supercritical carbon-dioxide extraction system with a 140 L autoclave and was first built in the 1980s and installed at KEK. After successful mass production of aerogels for the Belle experiment in the 1990s (see Section 4.1), KEK released both of the autoclaves around 2000 (the 31 L system for ethanol and 140 L system for carbon dioxide) and they were transferred to Chiba University in Chiba, Japan and the Mohri Oil Mill Co., Ltd., in Matsuzaka, Japan, respectively. The 140 L system is operated by the Mohri Oil Mill at the request of the JFCC (formerly Panasonic Electric Works) and the Chiba group. In addition, the third and relatively new supercritical carbon-dioxide extraction system with a 7.6 L autoclave was built at Chiba University in 2004 (see Figure 10). Since then, the Chiba group, led by the present author, has been developing modern aerogels. As shown in Figures 11 and 12, the pressure–temperature control profiles of each autoclave system were optimized to prevent the formation of cracks in the aerogel tiles.

The Chiba group developed a novel technique for producing aerogels with ultrahigh indices (ultrahigh densities), referred to as the pin-drying (previously, pinhole-drying) method [31]. The maximum index that can be attained in aerogels via the usual sol–gel synthesis method used by the Chiba group is approximately 1.14. Attempts to synthesize wet gels for aerogels with $n >$ 1.14 are hindered by the inhibition of the gelation by the excess amount of the silica precursor relative to the total sol volume (i.e., lack of diluent/solvent). Instead, of the direct synthesis of wet gels for high-index aerogels, the wet gel can be synthesized as usual as if targeting an aerogels with $n \approx 1.06$ and subsequently densified by partially evaporating the diluent solvent from the wet gel while maintaining the silica content (i.e., partially drying the wet gel) under ambient conditions. This is achieved by enclosing the synthesized wet gel in a special semi-sealed container with pinholes (called a pin container) as shown in Figure 13. The pin container enables the fabrication of silica-densified wet gels without cracking via the homothetic shrinkage of the wet gel (see Figure 14). Note that the pin-dried wet gel should be supercritically dried after the hydrophobic treatment to obtain the final aerogel; i.e., the pin-drying method is *not* a new drying (solvent-extraction) technique. The Chiba group employs the pin-drying method to fabricate dense aerogels with $n > 1.10$ and can produce aerogels with indices as high as $n = 1.26$ [31].

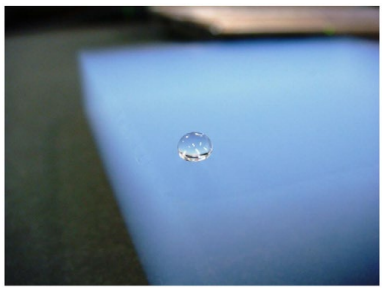

Figure 8. Water droplet on the surface of a hydrophobic aerogel tile with $n = 1.009$ fabricated at Chiba University (reproduced from [25] ©Elsevier B.V.).

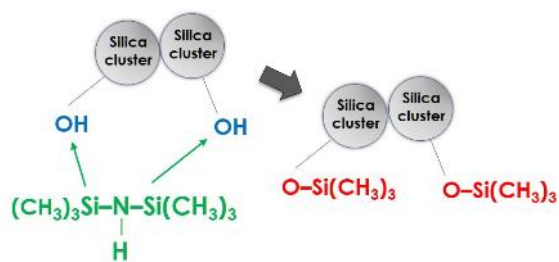

Figure 9. Trimethylsilyl [TMS, ―Si(CH$_3$)$_3$] modification of the surfaces of the silica nanoparticle clusters composing aerogels using HMDS.

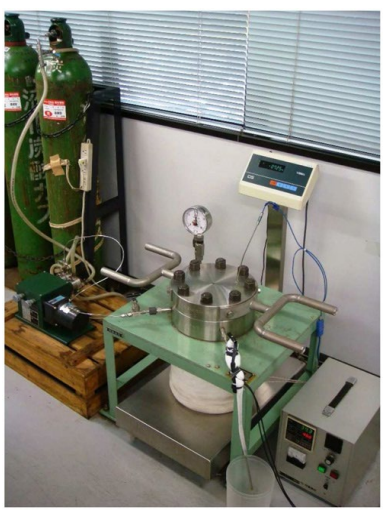

Figure 10. Custom-built supercritical carbon-dioxide extraction equipment with the 7.6 L autoclave installed at Chiba University (reproduced from [25] ©Elsevier B.V.).

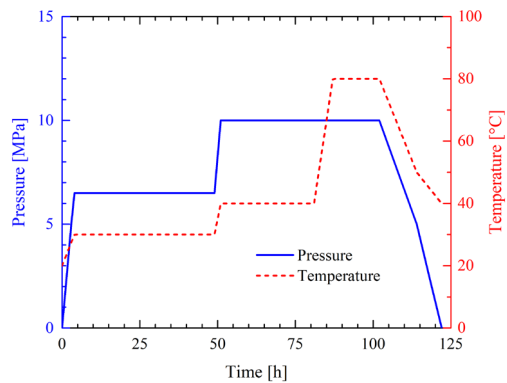

Figure 11. Designed pressure–temperature processes for supercritical carbon dioxide drying by modified Chiba pattern [28] adapted from the conventional KEK–Panasonic pattern [4] for the 140 L autoclave at Mohri Oil Mill (adapted from [29] ©Elsevier B.V.).

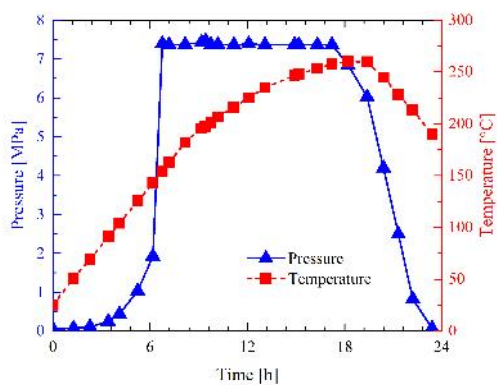

Figure 12. Measured pressure–temperature process of supercritical ethanol drying in the 31 L autoclave at Chiba University (adapted from [30] ©The Japanese Society for Biological Sciences in Space).

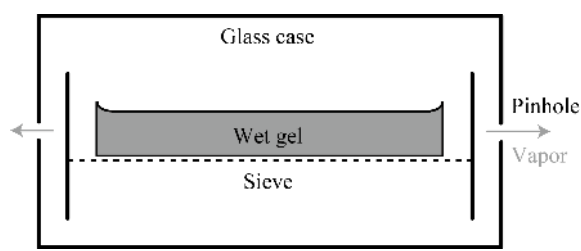

Figure 13. Schematic cross-sectional view of the first-generation pin container (adapted from [31] ©Elsevier B.V.).

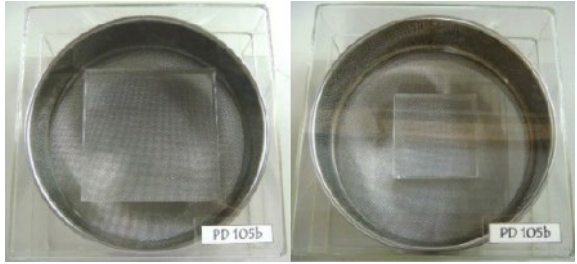

Figure 14. Top views of a wet gel at the beginning (left) and end (right) of pin drying (photos reproduced from [31] ©Elsevier B.V.)

3.2. Optical and Geometrical Characterization for Use as Cherenkov Radiators

Aerogels are one of the optical elements in Cherenkov counters and the optical parameters that are most critical for applications as Cherenkov radiators are the index and transparency. Figure 15 shows an aerogel's optical parameters as measured by the Chiba group [32] (refer to this plot for the following subsections). The geometric parameters, such as the tile shape and size (i.e., its volume, area, and thickness) are also important. In general, RICH counters have stricter requirements for the optical and geometric characteristics of aerogel radiators compared to threshold-type counters. In fact, RICH counters yielded practical applications, in part, through the improvement of the aerogel quality.

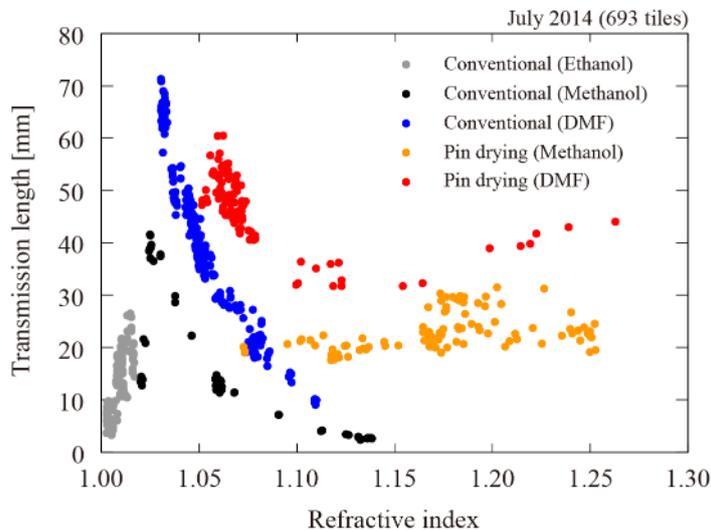

Figure 15. Transmission length at a wavelength of 400 nm as a function of the index for typical hydrophobic aerogels fabricated by the Chiba group between 2004 and 2014. The index was measured at a wavelength of 405 nm. Each data point represents an individual aerogel tile. Aerogels fabricated using the usual sol–gel synthesis method are classified by the diluent solvent used during the wet-gel synthesis: ethanol (gray), methanol (black), and DMF (blue). Aerogels produced by the pin-drying method are also classified according to the solvent used: methanol (orange) and DMF (red) (reproduced from [32] ©The Physical Society of Japan)

3.2.1. Refractive Index

The intermediate indices of aerogels between those of gases and condensed materials (i.e., liquids and solids) make aerogels highly useful as Cherenkov radiators. The typical index of aerogels with a certain level of transparency was previously known to range from approximately 1.01 to 1.05. Using the Cherenkov effect, such indices allow physicists to measure the velocity and identify types of charged subatomic particles at the momentum range from several hundred MeV/$c$ to multi-GeV/$c$, even in cases where velocity measurements cannot be performed using by time-of-flight counters due to the limitation of the time resolution. Moreover, aerogel indices can be tuned over a wide range by altering the production process, which is not the case with other condensed materials. The indices of gaseous radiators can be tuned by pressurizing or depressurizing them but this only produces a limited range of indices and requires the use of pressure vessels. Thus, the index is the most unique important characteristics of aerogels for use in Cherenkov radiators.

The index (in air) of an aerogel tile can be measured using a laser according to the minimum-deviation method based on the prism formula:

$$n = \sin\left(\frac{\alpha + \delta_{\min}}{2}\right)\left[\frac{1}{\sin(\alpha/2)}\right], (11)$$

where $\delta_{\min} = \tan^{-1}(d_{\min}/L)$ (see Figure 16 for the notation and an example of the measurement setup). The Chiba–KEK group manually rotates the aerogel table and visually reads the scale on a laser screen while a physicist group has implemented photodetectors to measure the location of the laser beam on the screen [33].

It is important to take into account the wavelength dependence of the aerogel index when designing Cherenkov counters. Previously, the (Chiba–)KEK group used a helium–neon (He–Ne) laser with $\lambda$ = 632.8 nm [24] because red light is clearly visible and readily penetrates aerogels (see Section 3.2.2). The Novosibirsk group also uses a red laser [34]. Since the 2000s, the Chiba–KEK group has used a blue–violet semiconductor laser with $\lambda$ = 405 nm [25] as well as a green He–Ne laser with $\lambda$ = 543.5 nm because typical photodetectors exhibit the maximum quantum efficiency around $\lambda \approx$ 400 nm. By measuring the indices at these three wavelengths, the chromatic dispersion of an aerogel can be measured [26].

In RICH counters, the absolute index value directly affects the Cherenkov angle according to Equation (4). An accurate measurement of the aerogel index is required for the precise determination of the particle velocity. Furthermore, the local index uniformity within a single monolithic aerogel tile is very important for RICH counters. Some RICH counters need to cover large solid angle around the particle-interaction point; thus, large-area, and/or large numbers of aerogel tiles are employed. It is generally preferred that the local index within a single monolithic tile is uniform within a variation of 1% depending on the requirements of the physics experiment and that the average indices of the segmented aerogel radiator tiles used in the RICH counter are identical to each other. One approach to improving the tile-uniformity is to enhance the quality control during aerogel production with respect to the methodology. The other approach is to fully characterize each tile's uniformity (i.e., to obtain the tile-uniformity map) prior to its installation or after installation using the event data from a calibration. To investigate the aerogel tile's index uniformity before installation, the Novosibirsk group uses a direct-measurement method based on the use of a laser [34] and the Chiba–KEK group developed an X-ray-based technique to indirectly measure the local index based on the density distribution in the aerogel [35].

In contrast, threshold-type Cherenkov counters do not require precise index values because the only condition is that the inequality in Equation (1) is fulfilled; in other words, only yes/no logic is required for PID. Of course, if the aerogel index is near the threshold index for the lower-velocity (AC = no) particles (i.e., if the aerogel index is sufficiently higher than the threshold index of the higher-velocity (AC = yes) particles), many Cherenkov photons will be emitted, which leads to a good counter performance. It should be noted that the aerogel's index is higher on the short-wavelength end of the visible light spectrum due to chromatic dispersion so if photodetectors have are sensitive to wavelengths in this range then the aerogel index may exceed the threshold index of the lower-velocity (AC = no) particles. However, this does not mean that the number of Cherenkov photons emitted immediately increase once the aerogel index reaches the threshold index.

The Chiba group can produce hydrophobic aerogels with any index between 1.0026 and 1.26 by adjusting the bulk density between 0.01 to 1.0 g/cm$^3$ [30]. Aerogels with indices between 1.0026 and 1.14 can be fabricated via the usual sol–gel synthesis method while those with $n$ = 1.14–1.26 must be produced by the pin-drying method. The Novosibirsk group reported that they can fabricate hygroscopic aerogels with $n$ = 1.006–1.13, with those having $n > 1.07$ being produced by a thermal-sintering technique [36]. The development of ultrahigh-index aerogels has enabled physicists to employ Cherenkov counters on the low-energy side and opened up new opportunities to conduct various high-energy physics experiments and, therefore, represents one of the greatest recent advances in the development of aerogels for Cherenkov counters.

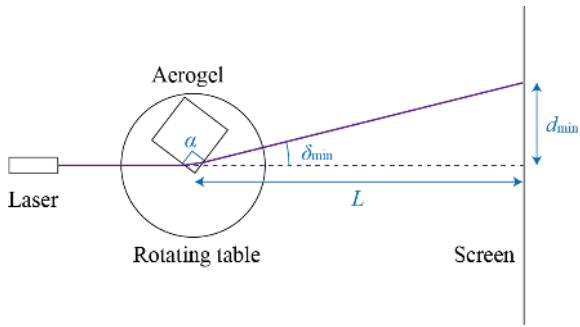

Figure 16. Setup used to conduct index measurements based on the minimum-deviation ($\delta_{min}$ angle) method. The corners (regarded simply as $\alpha = 90°$) of an aerogel tile are irradiated with a laser beam and the minimum laser deviation, $d_{min}$, is measured on a downstream screen separated by a distance of $L$ (adapted from [24] ©Elsevier B.V.).

3.2.2. Transparency

The optical transparency of an aerogel is another important factor that impacts its counter performance, especially when used in RICH counters. The transparencies of aerogels for use in Cherenkov radiators can be characterized in terms of the scattering and absorption lengths. Light propagation in aerogels is dominated by Rayleigh scattering so an aerogel's scattering length is considerably shorter than its absorption length. Only unscattered Cherenkov photons retain the Cherenkov angle information and, thus, can be used for Cherenkov imaging. Therefore, a long scattering length is vital for aerogels to be used in RICH counters. In contrast, the photon detection yield in threshold-type counters includes scattered photons that reach the photodetectors after multiple rounds of scattering and reflection on the inner walls of counter modules. Nevertheless, aerogels with short scattering lengths have a disadvantage in Cherenkov photon counting in that the light absorption by the aerogel cannot be ignored anymore.

The exact absorption length of an aerogel is difficult to determine experimentally. The Chiba–KEK group has only assumed that the absorption length is sufficiently longer than the scattering length such that the scattering length can be determined from the measured transmittance. The Novosibirsk group developed a method to determine the absorption length [34] and reported that the typical values are in 5–7 m at a wavelength of 400 nm in their aerogels [37]. This finding confirms that the absorption length is more than 100 times longer than the scattering length at 400 nm.

Considering that light absorption is negligible during transmission along the aerogel thickness of several centimeters, the scattering length, $L_{sc}$, is defined as

$$L_{sc}(\lambda) = -\frac{t}{\ln T(\lambda)}, \quad (12)$$

where $t$ is the thickness of the aerogel and $T$ is its transmittance as a function of the wavelength of light. The transmittance can be determined by measuring the ultraviolet–visible (UV–vis) spectrum using a spectrophotometer. In general, spectrophotometers use an integrating sphere to collect light from samples and the UV–vis spectrum strongly depends on the aerogel position relative to the light-integrating sphere. Thus, the Chiba–KEK group currently places aerogels 10 cm away from the entrance of the integrating sphere to minimize the amount of scattered light from the aerogel that is captured by the sphere as shown in Figure 17 [24]. The scattering length calculated using Equation (12) and the transmittance measured with this spectrophotometer configuration is referred to as the transmission length ($\Lambda_T$) by the Chiba–KEK group. The transmission length is usually measured at a wavelength of 400 nm because the photodetectors that are typically used in Cherenkov counters exhibit peak quantum efficiencies around that wavelength.

The UV–vis spectra of an aerogel obeys Rayleigh's inverse forth-power law:

$$T(\lambda, t) = A \exp(-Ct/\lambda^4), \quad (13)$$

where $A$ is the amplitude and $C$ is the clarity coefficient and is usually measured in units of $\mu m^4/cm$. Figure 18 shows a plot of the transmittance curve of a typical aerogel tile from the Chiba–KEK group that was obtained by their transmittance-measurement methodology. In this hydrophobic sample, the clarity coefficient was determined to be $C \approx 0.0053\ \mu m^4/cm$, corresponding to $\Lambda_T \approx 50$ mm. As shown in Figure 15, the transmission length depends on the index as well as the diluent/solvent used in the sol–gel synthesis of the wet gel. The Novosibirsk group also reported a similar clarity coefficient ($C \approx 0.0050\ \mu m^4/cm$) and measured scattering lengths of $L_{sc} = 5$–7 cm at 400 nm in hydrophilic aerogels with a wide range of indices [20,37]. However, the scattering (transmission) lengths measured by the two groups cannot be compared without taking into account the differences between the transmittance-measurement setups (i.e., the configurations of the aerogel sample and the light-integrating sphere) as discussed above.

The pin-drying technique facilitates the production of not only ultrahigh-index aerogels but also highly transparent aerogels. Pin-dried aerogels are generally more transparent than the original non-shrunk aerogels (Figure 19) [30]. Shrinkage due to pin-drying would alter the configuration of the silica nanoparticle clusters and pores compared to that in the original wet gel, resulting in improved transparency as discussed in [38]. It was also reported that the pin-drying method can improve the transparencies of aerogels with $n$ = 1.05–1.07 comparing to aerogels with the same index that were produced by the conventional method [39].

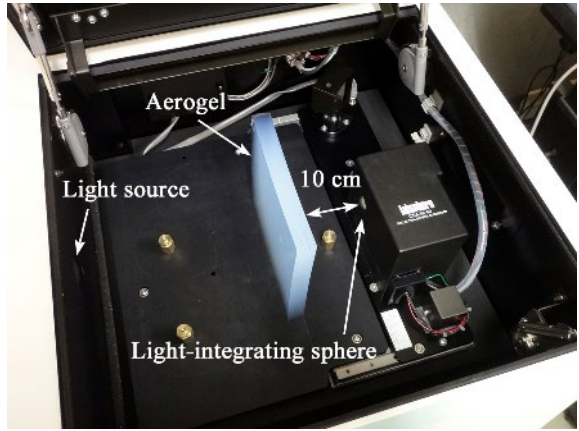

Figure 17. Measurement setup in the light-shielded chamber of a spectrophotometer (U-4100; Hitachi, Ltd., Japan) installed at KEK wherein the downstream surface of the aerogel is positioned 10 cm from the entrance of the light-integrating sphere (reproduced from [28] ©Elsevier B.V.).

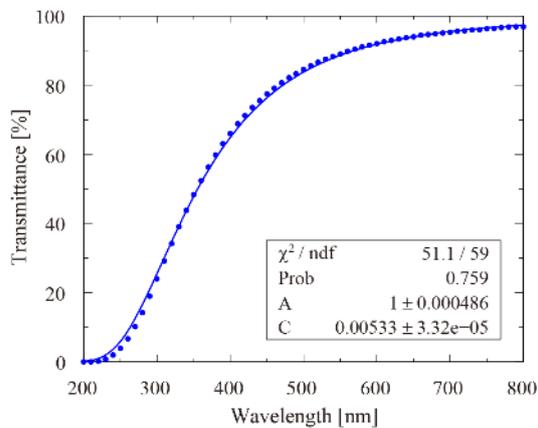

Figure 18. UV–vis spectrum for an aerogel with $n$ = 1.044 and $t$ = 20.8 mm fabricated at Chiba University. Circles show the transmittance values measured at intervals of 10 nm, and the solid line shows the curve fitted to this data based on Equation (13) having parameters of $A$ = 1 and $C$ = 0.00533 ± 0.00003 $\mu m^4$/cm (the upper limit of parameter $A$ was set to 1 in the fitting procedure). The corresponding transmission length was calculated to be approximately 50 mm at 400 nm (reproduced from [24] ©Elsevier B.V.).

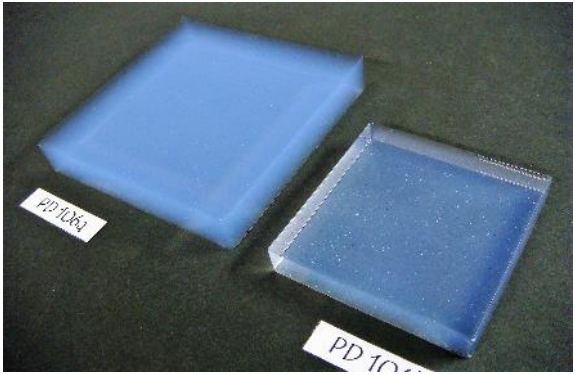

Figure 19. Pin-dried aerogel (right) with $n = 1.198$, $\Lambda_T = 39$ mm, and dimensions of 6.5 × 6.5 × 1 cm compared with the original non-shrunk aerogel (left) with $n = 1.067$ and $\Lambda_T = 32$ mm (adapted from [30] ©Elsevier B.V.).

### 3.2.3. Geometric Characteristics

The bulk density of fabricated aerogel tiles is gravimetrically measured as a geometric parameter based on the tile's volume and weight. The tile's thickness is used for calculating its scattering length in combination with its transmittance. It is known that the relationship between the index and density, $\rho$, of an aerogel is as follows [40]:

$$n^2 - 1 = \alpha\rho, (14)$$

where $\alpha$ is a proportionality constant. Empirically, Equation (14) can be approximated as a proportional relationship between the value of $n - 1$ and the density for small $n - 1$ values:

$$n - 1 = k\rho. (16)$$

Specifically, $n$ and the constant $k$ depend on the wavelength of the light. Figure 20 shows the relationship between the (absolute) index measured at a wavelength of 405 nm and the density in typical aerogels fabricated by the Chiba group. The results show that $k = 0.25$–$0.30$ cm$^3$/g depending on the diluent solvent used in the wet-gel synthesis [24]. The Novosibirsk group reported that $k \approx 0.21$ cm$^3$/g [40]. The difference between the values reported by the two groups is likely due to the presence or absence of surface modifications on the aerogels.

Researchers have been developing a technique for manufacturing large-area/large-volume aerogel tiles or blocks because Cherenkov counters are often used as subsystems in large-scale particle-spectrometer systems that also require large-area/large-volume aerogel radiator modules. Aerogels for use as Cherenkov radiators are usually manufactured in the shape of square monolithic tiles or blocks. Square aerogels allow the index to be measured at its four corners using lasers. Tile-shaped aerogels can be stacked along the thickness (i.e., particle-trajectory) direction and arranged into radiator modules. The Novosibirsk group previously reported they could produce aerogel blocks with dimensions of 20 × 20 × 5 cm with $n = 1.03$ [37]. The Chiba–KEK group succeeded in manufacturing high-index ($n \approx 1.05$), hydrophobic aerogel tiles with dimensions of 18 × 18 × 2 cm (Figure 21) with a low frequency of cracks (below 20%) [28]. Recently, the Novosibirsk group is focusing on fabricating large-area (hydrophilic) $n = 1.05$ aerogel tiles with dimensions of 20 × 20 × 2–3 cm [42].

It is usually difficult to machine aerogels once they have been fabricated and, thus, aerogel shapes are generally determined by the mold shape used in the wet-gel synthesis. However, hydrophobic characteristics allow aerogel tiles to be trimmed using a computer-automated water-jet-based cutting device without impacting their optical properties [4,28]. Water-jet machining is known as the only way to precisely cut aerogels, making full use of their hydrophobic characteristics. However, researchers must note that the cutting plane strongly scatters light. In aerogels that are to be used as manufactured, the molds must be designed by taking into account the wet-gel shrinkage during fabrication. The Novosibirsk group uses this approach because their aerogels are largely hydrophilic [42]. A fascinating molding technique for trapezoidal aerogels developed by the Chiba group is described in [43].

The integrity of tile shapes (e.g., with no bending) and the tile thickness uniformity are also important characteristics of aerogels to be used as radiators, especially RICH counters. The tilt of the downstream surface (the exit boundary of the Cherenkov light) in the tile affects the Cherenkov angle measurement as the boundary between the aerogels and the expansion gap before the photodetector (filled with gases such as air in most cases) refracts Cherenkov light. In other words, tilts in the tiles cause deviations in the trajectories of the Cherenkov photons. The Chiba group measured the thickness variation across a tile (more specifically, the outline of a cross-section of a tile) using a measuring microscope; however, this measurement requires the cross-section to be exposed by cutting the tile with the water-jet device [35]. Large-area tiles have to be handled carefully because they sometimes bend slightly. Furthermore, the top surface of the aerogel tile develops a meniscus geometry during the sol–gel synthesis in open-top molds, which causes a thickness variation across the tile. One solution is to place the tile's bottom surface (in contact with the mold in the sol–gel process) on the downstream side. Of course, the precise alignment of aerogel tiles in radiator modules is another important factor for the Cherenkov angle determination.

Multilayer monolithic aerogel blocks comprising different index layers can be produced by stacking sols for aerogels with different densities in a single mold during the wet-gel synthesis step. In the 1990s (or earlier), this concept was first developed for low-density aerogels used to capture intact cosmic dust from space at the Jet Propulsion Laboratory, California, U.S.A. [44,45]. As an ideal media for capturing intact cosmic dust, these multilayer aerogels were extended using a sol-pumping technique as a boundaryless aerogel with a density gradient from 0.01 to 0.05 g/cm$^3$ across the 3-cm thickness profile in a single monolithic block [45,46]. The density gradient was leveraged to collect cometary dust samples on a U.S. return mission, Stardust, and the aerogels captured dust from the comet Wild2, which returned to Earth in 2006 [45,47]. Motivated by the multilayer-focusing aerogel radiator scheme, the Chiba–KEK and Novosibirsk groups independently rediscovered multilayer monolithic aerogel blocks with different indices (in the around range $n \approx 1.05$) around 2004 [27,48], as shown in Figure 22. Aiming at the perfect focusing aerogels, the Novosibirsk group produced aerogel blocks with continuous index gradients (ranging from 1.036 to 1.042) along the thickness [49] based on [46]. The Chiba group developed a stereoscopic dual-layer (referred to as box-framed) aerogel tiles with densities of 0.01 and 0.03 g/cm$^3$ for capturing cosmic dust in low-Earth orbits; these are now used in the International Space Station for Japan's astrobiology mission, Tanpopo [50]. Mono-density (0.03 g/cm$^3$) aerogels previously developed by the KEK–Panasonic/Matsushita group have also been used for capturing cosmic dust in low-Earth orbits [51,52].

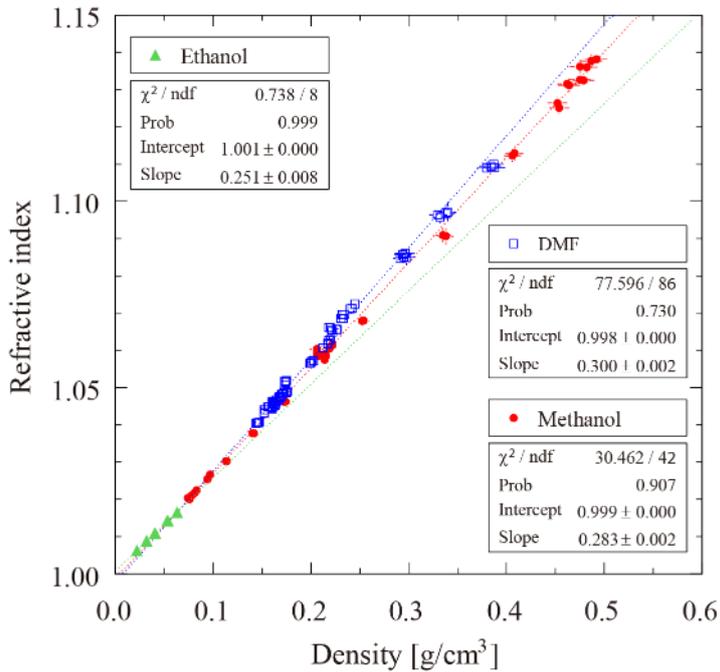

Figure 20. Relationship between the (absolute) index measured at 405 nm and the density in typical aerogels fabricated by the Chiba group. Each data point represents an individual aerogel tile. The aerogels fabricated using the usual sol–gel synthesis method were classified according to the diluent/solvent used during the wet-gel synthesis: ethanol (green triangles), methanol (red circles), and DMF (blue squares). The dotted lines represent best-fit linear functions to the three data sets; the corresponding *k* values are labeled as "slope" in the legends (reproduced from [24] ©Elsevier B.V.).

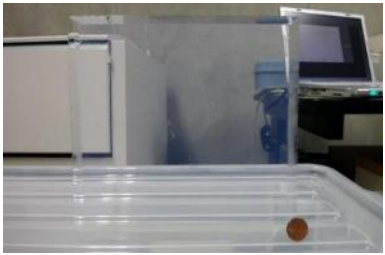

Figure 21. Large-area aerogel tile with dimensions of 18 × 18 × 2 cm designed by the Chiba–KEK group and manufactured by the JFCC. The refractive index and transmission length were 1.044 and 56 mm, respectively. This is the first tile delivered for use in the RICH detector of the Belle II experiment described in Section 4.2 (reproduced from [41] ©The authors, Tabata et al.).

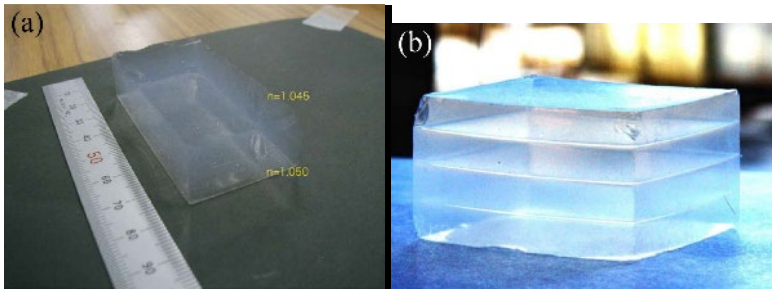

Figure 22. (a) Two-layer monolithic aerogel blocks produced by the Chiba–KEK group; the indices of the top and bottom layers are 1.045 and 1.050, respectively (reproduced from [27] ©Elsevier B.V.). (b) Four-layer monolithic aerogel block produced by the Novosibirsk group; where the indices range from 1.021 to 1.030 (reproduced from [48] ©Elsevier B.V.).

4. Applications to High-Energy-Physics Experiments

Among numerous high-energy-physics experiments that use AC counters, the B-factory experiment called Belle is the main focus in this section. The Belle experiment is a joint international research project involving the study of particle physics (mainly using *B*-meson systems) in Japan over the two generations. The Belle project was one of the most successful experiments using large-scale threshold-type aerogel Cherenkov counters, as described in detail in the first edition of this handbook [4]. The new Belle II experiment, which is almost ready to begin, is the successor of the previous Belle experiment and employs a state-of-the-art RICH detector system. In addition, several other experiments that use aerogels, including particle, nuclear, and cosmic-ray physics research projects, are presented herein.

4.1. B-Factory Experiment, Belle

The accelerator-based international project, Belle, was a particle physics experiment conducted at KEK during 1999–2010. The Belle experiment was performed using a 3 km-circumference circular electron–positron collider (called a KEKB collider) and a spectrometer (the Belle detector). The KEKB collider was operated at the energy determined to efficiently create *B* meson and anti-*B* meson pairs containing *b* quarks by electron–positron annihilation (i.e., 8 GeV electron and 3.5 GeV positron beams). The Belle detector was installed around the point of collision of the electron and positron beams to investigate the *B* meson decay products in detail. The KEKB facility was called the B-factory because it provided a large number of *B* mesons; thus, physicists could obtain statistically significant data about the *B* meson decay modes. The Belle spectrometer comprised several subsystems including a decay-vertex-position detector, a particle-tracking chamber with a superconducting magnet for momentum analysis, two kinds of PID counters, two kinds of electromagnetic calorimeters, and a neutral-particle detector [53].

The primary objective of the Belle experiment was to find the charge–parity (*CP*) symmetry violation in the *B* meson system, which was finally achieved in 2001. That led to the verification of the Kobayashi–Maskawa theory [54], which has since become a part of the Standard Model in particle physics. The Kobayashi–Maskawa theory offers insight into the solution of a major question in physics: "Why the Universe is dominated by not anti-matter but matter?" M.

Kobayashi and T. Maskawa shared the Nobel Prize in Physics 2008 (with Y. Nambu) "for the discovery of the origin of the broken symmetry which predicts the existence of at least three families of quarks in nature" [55]. The major achievements in physics due to the Belle experiment are summarized briefly in [56] and the most fruitful findings were fully reviewed in a paper titled *The Physics of the B factories*, which published by the Belle collaboration and the other B-factory experiment collaboration in the U.S.A. [57].

One of the two PID counters in the Belle detector was a threshold-type Aerogel Cherenkov Counter (ACC) [58–60]. The Belle ACC was composed of a total of 1188 counter modules installed around the collision point of the beams and was mainly used to separate kaons from pions at a few GeV/*c*. The counter configurations were slightly different depending on the position relative to the collision point but the counter boxes were basically lined with diffusive reflector sheets and equipped with one or two 2–3 in diameter PMTs that can be operated in the strong magnetic field (1.5 T) of the Belle spectrometer. Typical dimensions of the counter boxes were approximately $12 \times 12 \times 12$ cm, which was filled with five aerogel tiles having $n = 1.01$–1.03. The ACC modules were designed after prototyping AC counters and evaluating their performances using test beams [23].

In the 1990s, low-index, hydrophobic aerogel tiles with $n = 1.01$–1.03 were mass-produced using the 140 L autoclave system by the KEK–Matsushita group at KEK [58–60]. A total of approximately 2 $m^3$ of high-transparency aerogels were manufactured in-house. The absence of a significant degradation in the counter's light yield over the 10-year ACC operation period demonstrated the integrity of the hydrophobic characteristics of the aerogels fabricated at KEK. This successful application of Japanese hydrophobic aerogels would greatly contribute to the successful commercialization of Panasonic/Matsushita aerogels.

4.2. Super-B-Factory Experiment, Belle II

To explore physics beyond the Standard Model of particle physics by precisely investigating the decay processes of particles such as *B* mesons and *τ* leptons, the super-B-factory experiment is planned at KEK. The new experiment, Belle II [61], is going to be conducted using an upgraded SuperKEKB collider and the Belle II detector (Figure 23), which are successors of the previous KEKB collider and Belle detector, respectively; these components are now in the final phase of commissioning. The Belle II collaboration comprised over 700 members of 108 institutes in 25 countries and regions. The SuperKEKB collider, accelerating 7 GeV electron and 4 GeV positron beams, is designed to surpass (by a factor of 40) its own world record for instantaneous luminosity ($2.11 \times 10^{34}$ cm$^{-2}$ s$^{-1}$ was achieved by the previous KEKB accelerator); thus, it is called a super-B-factory.

One of the major detector upgrades is the transition from the Belle ACC to the Aerogel RICH (ARICH) detector. The ARICH detector was positioned in the direction of the electron beam travel in the Belle II detector, which is called the forward end cap, in late 2017. The cylindrically shaped forward end cap has a diameter of more than 2 m (Figure 24); however, the dimension along its thickness direction is limited to approximately 30 cm. This resulted in the need for a proximity-focusing RICH counter. The ARICH detector comprises two support structure modules for aerogel radiators and photodetectors. The aerogel radiator module, which was designed based on the results from the study of large-area aerogel tiles [28], is shown in

Figure 25. A total of 420 hybrid avalanche photodetectors pixelated to 144 channels were used as a Cherenkov light sensor [63].

In 2004, the Belle ARICH group first proposed a novel focusing multilayer aerogel radiator in which the layers have different indices [64]. Considering dual-layer configuration (Figure 26), the index of the downstream layer must be higher than that of the upstream layer and be chosen precisely so that the Cherenkov ring images from the two layers overlap on the photodetection plane. Note that the Cherenkov angle in the downstream layer is larger than that in the upstream layer based on Equation (4). This scheme can increase the number of detected photoelectrons by increasing the total thickness of the aerogel tiles while keeping a high Cherenkov angle resolution for a single photon. The same concept was presented by the Novosibirsk group around the same time [48]. With the help of this scheme and the choice of aerogels with $n =$ 1.045 and 1.055, the ARICH detector enables precise separation of kaons from pions at high momenta up to 4 GeV/$c$, under which conditions the previous end-cap ACC could not perform PID.

The dual-layer radiator scheme together with the tiling scheme described above requires a total of 248 aerogel tiles [11]. Thus, the Chiba–KEK researchers in collaboration with Panasonic and the JFCC developed a method for producing 18 × 18 × 2 cm tiles while maintaining a high crack-free yield [28]. The 140 L autoclave system was used again to mass-produce approximately 450 high-index ($n$ = 1.045 and 1.055) tiles for the ARICH detector during 2013–2014 (refer back to Figure 21). Crack-free tiles were then trimmed into wedge (fan) shapes with a water-jet cutter. By the end of 2016, all tiles were mounted in the radiator module (Figure 27) [65] and the ARICH subsystem was installed in the Belle II detector in late 2017. The first collision between the electron and positron beams is expected to occur in April 2018 or later. Physicists believe that the Belle II experiment will lead to the discovery of new physics phenomena beyond our expectations within the next 10 years.

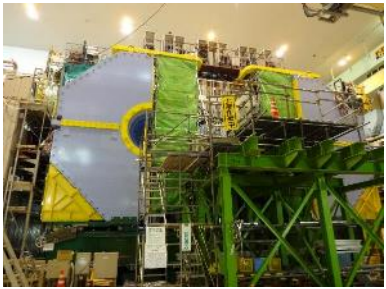

Figure 23. Belle II detector during the upgrade process at the roll-out position from the SuperKEKB beam line (as of December 2016). The detector measures approximately 8 × 8 × 8 m and 1400 ton.

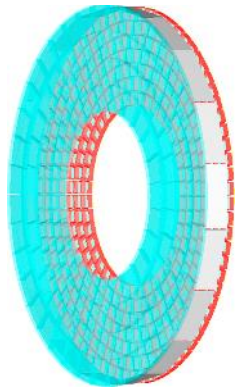

Figure 24. Drawing of the ARICH detector by a detector simulation software. The blue and red discs show the aerogel radiator and photodetector modules, respectively (reproduced from [62] ©Elsevier B.V.).

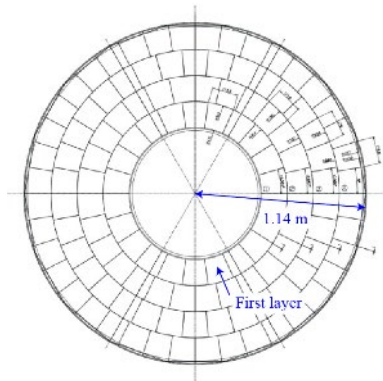

Figure 25. CAD drawing of the ARICH radiator tiling in the cylindrical forward end cap of the Belle II detector with an outer radius of 1.14 m and a tiling area comprising four concentric layers and 124 segments for installing trimmed aerogel tiles (reproduced from [28] ©Elsevier B.V.).

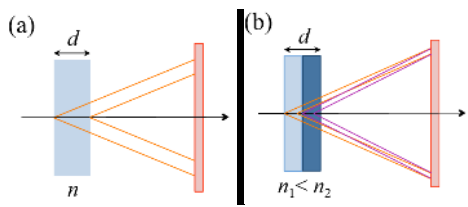

Figure 26. Schematics of (a) the basic proximity-focusing and (b) the focusing dual-layer aerogel RICH counters (adapted from [16] ©The authors, Iwata et al.).

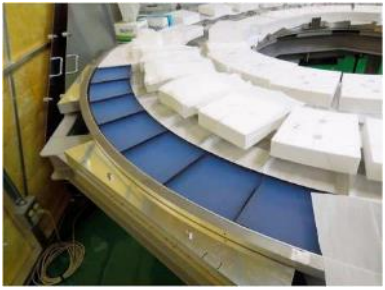

Figure 27. Wedge-shaped aerogel tiles installed in the radiator module of the ARICH detector (adapted from [65] ©Elsevier B.V.).

4.3. Other Experiments

Aerogels produced by the Novosibirsk group are used or are expected to be used in the RICH counters of several ongoing and planned experiments. The LHCb experiment [66] that is being conducted using the Large Hadron Collider (LHC) at the European Organization for Nuclear Research (CERN) in Switzerland is investigating *b*-physics similar to the KEK (super-)B-factory. The AMS-02 experiment [67] that is being performed aboard the International Space Station to observe primary cosmic rays and study dark matter physics, antimatter physics, etc. The CLAS12 experiment [45] is a planned accelerator-based nuclear-physics program at the Thomas Jefferson National Accelerator Facility in Virginia, U.S.A.

Aerogels developed by the Chiba group are also used in several experiments. In the Japan Proton Accelerator Research Complex (J-PARC) in Japan, various particle- and nuclear-physics experiments were conducted, are being conducted, or are planned. Many of these experiments employ threshold-type Cherenkov counters with aerogels having different indices. For example, the J-PARC E36 experiment [42], which is a particle physics program to test lepton flavor universality, was performed in 2015 with AC counters containing the $n = 1.08$ aerogels shown in Figure 4. Recently, aerogels made by the Chiba group were used in an AC counter for a nuclear-physics experiment conducted at the GSI Helmholtz Centre for Heavy Ion Research in Germany. [68] The Chiba group is currently developing aerogels for use in the RICH counters for an Antarctic balloon-borne cosmic-ray physics experiment [69] and a hadron production measurement program to be performed at the Fermi National Accelerator Laboratory in Illinois, U.S.A.

5. Summary

In this chapter, we reviewed the application of silica aerogels as Cherenkov radiators in high-energy physics experiments. Aerogels have outstanding optical properties: they have high transparency and tunable refractive indices that fill the gap between those of gases and condensed materials. Thus, they are useful and promising media for use in Cherenkov counters. We reviewed physics of Cherenkov radiation and the operational principles of threshold-type Cherenkov counters and RICH counters. Aerogel-production technologies and achievements toward improved counter performance were described along with a history of the activities of the two major groups working on AC counters, the Russian and Japanese groups. We discussed the optical and geometric properties of aerogels, ways to characterize these properties, and the requirements for their use as Cherenkov radiators. As specific applications, the threshold-type ACC for the first-generation B-factory experiment was reviewed and the novel ARICH counter that is now in the final phase of commissioning for the next-generation super B-factory experiment was described. Additional applications in other experiments were also briefly summarized. Currently, aerogels are actively used as Cherenkov radiators all over the world and physicists plan to make full use of the recently developed high-quality aerogels in a wide variety of experiments in future.

Acknowledgments

The author is grateful to Drs. H. Yokogawa and K. Kugimiya of Panasonic Corporation for giving me the opportunity to write this chapter. I am also grateful to Prof. H. Kawai of Chiba University, Prof. I. Adachi of KEK, and Prof. T. Sumiyoshi of Tokyo Metropolitan University, Japan for their support to the aerogel development at Chiba University. This chapter is the result of many fertile discussions with them.

References

1. S. S. Kistler: Coherent expanded aerogels and jellies, Nature **127**, 741 (1931)

2. M. Cantin, M. Casse, L. Koch, R. Jouan, P. Mestreau, D. Roussel, F. Bonnin, J. Moutel, S. J. Teichner: Silica aerogels used as Cherenkov radiators, Nucl. Instrum. Methods **118**, 177–182 (1974)

3. M. Tabata, Chem. Chem. Ind. **70**, 116–118 (2017) (the tiles is in Japanese)

4. H. Yokogawa: Transparent silica aerogel blocks for high-energy physics research. In: *Aerogels Handbook*, ed. by M. A. Aegerter, N. Leventis, M. M. Koebel (Springer New York, New York, NY 2011), pp. 651–663

5. P. A. Čerenkov: Visible luminescence of pure liquids under the influence of γ-radiation, Dokl. Akad. Nauk SSSR **2**, 451–454 (1934)

6. P. A. Čerenkov: Visible radiation produced by electrons moving in a medium with velocities exceeding that of light, Phys. Rev. **52**, 378–379 (1937)

7. I. M. Frank, I. E. Tamm: Coherent visible radiation of fast electrons passing through matter, Dokl. Akad. Nauk SSSR **14**, 107–112 (1937)

8. I. Frank, I. Tamm: Coherent visible radiation of fast electrons passing through matter. In: *Selected Papers*, doi:10.1007/978-3-642-74626-0_2, ed. by B. M. Bolotovskii, V. Y. Frenkel, R. Peierls (Springer Berlin Heidelberg, Berlin, Heidelberg 1991), pp. 29–35

9. Nobel Media AB: The Nobel Prize in Physics 1958, Nobelprize.org, https://www.nobelprize.org/nobel_prizes/physics/laureates/1958/ (2014)

10. C. Patrignani et al. (Particle Data Group): Review of particle physics, Chin. Phys. C **40**, 100001 (2016)

11. M. Tabata, I. Adachi, N. Hamada, K. Hara, T. Iijima, S. Iwata, H. Kakuno, H. Kawai, S. Korpar, P. Križan, T. Kumita, S. Nishida, S. Ogawa, R. Pestotnik, L. Šantelj, A. Seljak, T. Sumiyoshi, E. Tahirović, K. Yoshida, Y. Yusa: Silica aerogel radiator for use in the A-RICH system utilized in the Belle II experiment, Nucl. Instrum. Methods Phys. Res. A **766**, 212–216 (2014)

12. R. Pestotnik, I. Adachi, R. Dolenec, K. Hataya, S. Iori, S. Iwata, H. Kakuno, R. Kataura, H. Kawai, H. Kindo, T. Kobayashi, S. Korpar, P. Križan, T. Kumita, M. Mrvar, S. Nishida, K. Ogawa, S. Ogawa, L. Šantelj, T. Sumiyoshi, M. Tabata, M. Yonenaga, Y. Yusa: The aerogel ring imaging Cherenkov system at the Belle II spectrometer, Nucl. Instrum. Methods Phys. Res. A **876**, 265–268 (2017)

13. K. A. Olive et al. (Particle Data Group): Review of particle physics, Chin. Phys. C **38**, 090001 (2014)

14. T. Iijima, S. Korpar, I. Adachi, S. Fratina, T. Fukushima, A. Gorišek, H. Kawai, H. Konishi, Y. Kozakai, P. Križan, T. Matsumoto, Y. Mazuka, S. Nishida, S. Ogawa, S. Ohtake, R. Pestotnik, S. Saitoh, T. Seki, T. Sumiyoshi, Y. Uchida, Y. Unno, S. Yamamoto: A novel type of proximity focusing RICH counter with multiple refractive index aerogel radiator, Nucl. Instrum. Methods Phys. Res. A **548**, 383–390 (2005)

15. R. Pestotnik, P. Križan, S. Korpar, T. Iijima: Design optimization of the proximity focusing RICH with dual aerogel radiator using a maximum-likelihood analysis of Cherenkov rings, Nucl. Instrum. Methods Phys. Res. A **595**, 256–259 (2008)

16. S. Iwata, I. Adachi, K. Hara, T. Iijima, H. Ikeda, H. Kakuno, H. Kawai, T. Kawasaki, S. Korpar, P. Križan, T. Kumita, S. Nishida, S. Ogawa, R. Pestotnik, L. Šantelj, A. Seljak, T. Sumiyoshi, M. Tabata, E. Tahirović, Y. Yusa: Particle identification performance of the prototype aerogel RICH counter for the Belle II experiment, Progr. Theor. Exp. Phys. **2016**, 033H001 (2016)

17. A. F. Danilyuk, E. A. Kravchenko, A. G. Okunev, A. P. Onuchin, S. A. Shaurman: Synthesis of aerogel tiles with high light scattering length, Nucl. Instrum. Methods Phys. Res. A **433**, 406–407 (1999)

18. H. Kawai, J. Haba, T. Homma, M. Kobayashi, K. Miyake, T. S. Nakamura, N. Sasao, Y. Sugimoto, M. Yoshioka, M. Daigo: Tests of a silica aerogel Cherenkov counter, Nucl. Instrum. Methods Phys. Res. A **228**, 314–322 (1985)

19. A. Buzykaev, A. Danilyuk, S. Ganzhur, T. Gorodetskaya, E. Kravchenko, A. Onuchin, A. Vorobiov: Aerogels with high optical parameters for Cherenkov counters, Nucl. Instrum. Methods Phys. Res. A **379**, 465–467 (1996)

20. A. Y. Barnyakov, M. Y. Barnyakov, J. Bähr, T. Bellunato, K. I. Beloborodov, V. S. Bobrovnikov, A. R. Buzykaev, M. Calvi, A. F. Danilyuk, V. Djordjadze, V. B. Golubev, S. A. Kononov, E. A. Kravchenko, D. Lipka, C. Matteuzzi, M. Musy, A. P. Onuchin, D. Perego, V. A. Rodiakin, G. A. Savinov, S. I. Serednyakov, A. G. Shamov, F. Stephan, V. A. Tayursky, A. I. Vorobiov: Development of aerogel Cherenkov detectors at Novosibirsk, Nucl. Instrum. Methods Phys. Res. A **553**, 125–129 (2005)

21. A. Y. Barnyakov, M. Y. Barnyakov, V. V. Barutkin, V. S. Bobrovnikov, A. R. Buzykaev, A. F. Daniluk, S. A. Kononov, V. L. Kirillov, E. A. Kravchenko, A. P.

Onuchin: Influence of water on optical parameters of aerogel, Nucl. Instrum. Methods Phys. Res. A **598**, 166–168 (2009)

22. D. L. Perego: Ageing tests and recovery procedures of silica aerogel, Nucl. Instrum. Methods Phys. Res. A **595**, 224–227 (2008)

23. H. Yokogawa, M. Yokoyama: Hydrophobic silica aerogels, J. Non-Cryst. Solids **186**, 23–29 (1995)

24. I. Adachi, T. Sumiyoshi, K. Hayashi, N. Iida, R. Enomoto, K. Tsukada, R. Suda, S. Matsumoto, K. Natori, M. Yokoyama, H. Yokogawa: Study of a threshold Cherenkov counter based on silica aerogels with low refractive indices, Nucl. Instrum. Methods Phys. Res. A **355**, 390–398 (1995)

25. M. Tabata, I. Adachi, H. Kawai, T. Sumiyoshi, H. Yokogawa: Hydrophobic silica aerogel production at KEK, Nucl. Instrum. Methods Phys. Res. A **668**, 64–70 (2012)

26. M. F. Villoro, J. C. Plascencia, R. Núñez, A. Menchaca-Rocha, J. M. Hernández, E. Camarillo, M. Buénerd: Measurement of the dispersion law for hydrophobic silica aerogel SP-25, Nucl. Instrum. Methods Phys. Res. A **480**, 456–462 (2002)

27. Japan Fine Ceramics Center: http://www.jfcc.or.jp/en/

28. I. Adachi, S. Fratina, T. Fukushima, A. Gorišek, T. Iijima, H. Kawai, M. Konishi, S. Korpar, Y. Kozakai, P. Križan, T. Matsumoto, Y. Mazuka, S. Nishida, S. Ogawa, S. Ohtake, R. Pestotnik, S. Saitoh, T. Seki, T. Sumiyoshi, M. Tabata, Y. Uchida, Y. Unno, S. Yamamoto: Study of highly transparent silica aerogel as a RICH radiator, Nucl. Instrum. Methods Phys. Res. A **553**, 146–151 (2005)

29. M. Tabata, I. Adachi, Y. Hatakeyama, H. Kawai, T. Morita, T. Sumiyoshi: Large-area silica aerogel for use as Cherenkov radiators with high refractive index, developed by supercritical carbon dioxide drying, J. Supercrit. Fluids **110**, 183–192 (2016)

30. M. Tabata, Y. Kawaguchi, S. Yokobori, H. Kawai, J. Takahashi, H. Yano, A. Yamagishi: Tanpopo cosmic dust collector: Silica aerogel production and bacterial DNA contamination analysis, Biol. Sci. Space **25**, 7–12 (2011)

31. M. Tabata, I. Adachi, Y. Ishii, H. Kawai, T. Sumiyoshi, H. Yokogawa: Development of transparent silica aerogel over a wide range of densities, Nucl. Instrum. Methods Phys. Res. A **623**, 339–341 (2010)

32. M. Tabata, H. Kawai: Progress in development of silica aerogel for particle- and nuclear-physics experiments at J-PARC, JPS Conf. Proc. **8**, 022004 (2015)

33. Y. Sallaz-Damaz, L. Derome, M. Mangin-Brinet, M. Loth, K. Protasov, A. Putze, M. Vargas-Trevino, O. Véziant, M. Buénerd, A. Menchaca-Rocha, E. Belmont, M. Vargas-Magaña, H. Léon-Vargas, A. Ortiz-Velàsquez, A. Malinine, F. Barão, R. Pereira, T. Bellunato, C. Matteuzzi, D. L. Perego: Characterization study of silica aerogel for Cherenkov imaging, Nucl. Instrum. Methods Phys. Res. A **614**, 184–195 (2010)

34. A. R. Buzykaev, A. F. Danilyuk, S. F. Ganzhur, E. A. Kravchenko, A. P. Onuchin: Measurement of optical parameters of aerogel, Nucl. Instrum. Methods Phys. Res. A **433**, 396–400 (1999)

35. M. Tabata, Y. Hatakeyama, I. Adachi, T. Morita, K. Nishikawa: X-ray radiographic technique for measuring density uniformity of silica aerogel, Nucl. Instrum. Methods Phys. Res. A **697**, 52–58 (2013)

36. A. Y. Barnyakov, M. Y. Barnyakov, K. I. Beloborodov, V. S. Bobrovnikov, A. R. Buzykaev, A. F. Danilyuk, V. B. Golubev, V. V. Gulevich, S. A. Kononov, E. A. Kravchenko, A. P. Onuchin, K. A. Martin, S. I. Serednyakov, V. M. Vesenev: Particle identification system based on dense aerogel, Nucl. Instrum. Methods Phys. Res. A **732**, 330–332 (2013)

37. A. Y. Barnyakov, M. Y. Barnyakov, K. I. Beloborodov, V. S. Bobrovnikov, A. R. Buzykaev, A. F. Danilyuk, V. B. Golubev, V. V. Gulevich, V. L. Kirillov, S. A. Kononov, E. A. Kravchenko, A. P. Onuchin, S. I. Serednyakov: Results from R&D of Cherenkov detectors at Novosibirsk, Nucl. Instrum. Methods Phys. Res. A **581**, 410–414 (2007)

38. T. Li, B. Zhou, A. Du, Y. Xiang, S. Wu, M. Liu, W. Ding, J. Shen, Z. Zhang: Microstructure control of the silica aerogels via pinhole drying, J. Sol–Gel Sci. Technol. **84**, 96–103 (2017)

39. M. Tabata, I. Adachi, Y. Hatakeyama, H. Kawai, T. Morita, K. Nishikawa: Optical and radiographical characterization of silica aerogel for Cherenkov radiator, IEEE Trans. Nucl. Sci. **59**, 2506–2511 (2012)

40. A. F. Danilyuk, V. L. Kirillov, M. D. Savelieva, V. S. Bobrovnikov, A. R. Buzykaev, E. A. Kravchenko, A. V. Lavrov, A. P. Onuchin: Recent results on aerogel development for use in Cherenkov counters, Nucl. Instrum. Methods Phys. Res. A **494**, 491–494 (2002)

41. M. Tabata, I. Adachi, H. Kawai, S. Nishida, T. Sumiyoshi: Recent progress in the development of large area silica aerogel for use as RICH Radiator in the Belle II experiment, In: Proc. 3rd Int'l Conf. Technol. Instrum. Part. Phys., (2014) PoS(TIPP2014)327

42. M. Contalbrigo, I. Balossino, L. Barion, A. Y. Barnyakov, G. Battaglia, A. F. Danilyuk, A. A. Katcin, E. A. Kravchenko, M. Mirazita, A. Movsisyan, D. Orecchini, L. L. Pappalardo, S. Squerzanti, S. Tomassini, M. Turisini: Aerogel Mass production for the CLAS12 RICH: Novel characterization methods and optical performance, Nucl. Instrum. Methods Phys. Res. A **876**, 168 168–172 (2017)

43. M. Tabata, A. Toyoda, H. Kawai, Y. Igarashi, J. Imazato, S. Shimizu, H. Yamazaki: Fabrication of silica aerogel with $n$=1.08 for $e^+/\mu^+$ separation in a threshold Cherenkov counter of the J-PARC TREK/E36 experiment, Nucl. Instrum. Methods Phys. Res. A **795**, 206–212 (2015)

44. P. Tsou: Silica aerogel captures cosmic dust intact, J. Non-Cryst. Solids **186**, 415–427 (1995)

45. P. Tsou, D. E. Brownlee, S. A. Sandford, F. Hörz, M. E. Zolensky: Wild 2 and interstellar sample collection and Earth return, J. Geophys. Res.: Planets **108**, 8113 (2003)

46. S. M. Jones: A method for producing gradient density aerogel, J. Sol–Gel Sci. Technol. **44**, 255–258 (2007)

47. D. Brownlee et al.: Comet 81p/Wild 2 under a microscope, Science **314**, 1711–1716 (2006)

48. A. Y. Barnyakov, M. Y. Barnyakov, V. S. Bobrovnikov, A. R. Buzykaev, A. F. Danilyuk, V. L. Kirillov, S. A. Kononov, E. A. Kravchenko, A. P. Onuchin: Focusing aerogel RICH (FARICH), Nucl. Instrum. Methods Phys. Res. A **553**, 70–75 (2005)

49. A. Y. Barnyakov, M. Y. Barnyakov, V. S. Bobrovnikov, A. R. Buzykaev, V. V. Gulevich, A. F. Danilyuk, S. A. Kononov, E. A. Kravchenko, I. A. Kuyanov, S. A. Lopatin, A. P. Onuchin, I. V. Ovtin, N. A. Podgornov, V. V. Porosev, A. Y. Predein, R. S. Protsenko: Aerogel for FARICH detector, Nucl. Instrum. Methods Phys. Res. A **766**, 235–236 (2014)

50. M. Tabata, H. Kawai, H. Yano, E. Imai, H. Hashimoto, S. Yokobori, A. Yamagishi: Ultralow-density double-layer silica aerogel fabrication for the intact capture of cosmic dust in low-Earth orbits, J. Sol–Gel Sci. Technol. **77**, 325–334 (2016)

51. Y. Kitazawa, A. Fujiwara, T. Kadono, K. Imagawa, Y. Okada, K. Uematsu: Hypervelocity impact experiments on aerogel dust collector, J. Geophys. Res.: Planets **104**, 22035–22052 (1999)

52. T. Noguchi, T. Nakamura, T. Ushikubo, N. T. Kita, J. W. Valley, R. Yamanaka, Y. Kimoto, Y. Kitazawa: A chondrule-like object captured by space-exposed aerogel on the International Space Station, Earth Planet. Sci. Lett. **309**, 198–206 (2011)

53. A. Abashian et al. (Belle Collaboration): The Belle detector, Nucl. Instrum. Methods Phys. Res. A **479**, 117–232 (2002)

54. M. Kobayashi, T. Maskawa: *CP*-violation in the renormalizable theory of weak interaction, Progr. Theor. Phys. **49**, 652–657 (1973)

55. Nobel Media AB: The Nobel Prize in Physics 2008, Nobelprize.org, https://www.nobelprize.org/nobel_prizes/physics/laureates/2008/ (2014)

56. J. Brodzicka, T. Browder, P. Chang, S. Eidelman, B. Golob, K. Hayasaka, H. Hayashii, T. Iijima, K. Inami, K. Kinoshita, Y. Kwon, K. Miyabayashi, G. Mohanty, M. Nakao, H. Nakazawa, S. Olsen, Y. Sakai, C. Schwanda, A. Schwartz, K. Trabelsi, S. Uehara, S. Uno, Y. Watanabe, A. Zupanc: Physics achievements from the Belle experiment, Progr. Theor. Exp. Phys. **2012**, 04D001 (2012)

57. A. J. Bevan et al.: The physics of the *B* factories, Eur. Phys. J. C **74**, 3026 (2014)

58. T. Sumiyoshi, I. Adachi, R. Enomoto, T. Iijima, R. Suda, M. Yokoyama, H. Yokogawa: Silica aerogels in high energy physics, J. Non-Cryst. Solids **225**, 369–374 (1998)

59. T. Sumiyoshi, I. Adachi, R. Enomoto, T. Iijima, R. Suda, C. Leonidopoulos, D. R. Marlow, E. Prebys, R. Kawabata, H. Kawai, T. Ooba, M. Nanao, K. Suzuki, S. Ogawa, A. Murakami, M. H. R. Khan: Silica aerogel Cherenkov counter for the KEK B-factory experiment, Nucl. Instrum. Methods Phys. Res. A **433**, 385–391 (1999)

60. T. Iijima, I. Adachi, R. Enomoto, R. Suda, T. Sumiyoshi, C. Leonidopoulos, D. R. Marlow, E. Prebys, H. Kawai, E. Kurihara, M. Nanao, K. Suzuki, Y. Unno, S. Ogawa, A. Murakami, M. H. R. Khan: Aerogel Cherenkov counter for the Belle detector, Nucl. Instrum. Methods Phys. Res. A **453**, 321–325 (2000)

61. Belle II Collaboration (Internet site provider, Deutsche Elektronen-Synchrotron DESY): https://www.belle2.org/

62. L. Šantelj, I. Adachi, R. Dolenec, K. Hataya, S. Iori, S. Iwata, H. Kakuno, R. Kataura, H. Kawai, H. Kindo, T. Kobayashi, S. Korpar, P. Križan, T. Kumita, M. Mrvar, S. Nishida, K. Ogawa, S. Ogawa, R. Pestotnik, T. Sumiyoshi, M. Tabata, M. Yonenaga, Y. Yusa: Recent developments in software for the Belle II aerogel RICH, Nucl. Instrum. Methods Phys. Res. A **876**, 104–107 (2017)

63. S. Nishida, I. Adachi, N. Hamada, K. Hara, T. Iijima, S. Iwata, H. Kakuno, H. Kawai, S. Korpar, P. Križan, S. Ogawa, R. Pestotnik, L. Šantelj, A. Seljak, T. Sumiyoshi, M. Tabata, E. Tahirović, K. Yoshida, Y. Yusa: Development of a 144-channel Hybrid Avalanche Photo-Detector for Belle II ring-imaging Cherenkov counter with an aerogel radiator, Nucl. Instrum. Methods Phys. Res. A **787**, 59–63 (2015)

64. T. Iijima, S. Korpar, I. Adachi, S. Fratina, T. Fukushima, A. Gorišek, H. Kawai, H. Konishi, Y. Kozakai, P. Križan, T. Matsumoto, Y. Mazuka, S. Nishida, S. Ogawa, S. Ohtake, R. Pestotnik, S. Saitoh, T. Seki, T. Sumiyoshi, Y. Uchida, Y. Unno, S. Yamamoto: A novel type of proximity focusing RICH counter with multiple refractive index aerogel radiator, Nucl. Instrum. Methods Phys. Res. A **548**, 383–390 (2005)

65. I. Adachi, R. Dolenec, K. Hataya, S. Iori, S. Iwata, H. Kakuno, R. Kataura, H. Kawai, H. Kindo, T. Kobayashi, S. Korpar, P. Križan, T. Kumita, M. Mrvar, S. Nishida, K. Ogawa, S. Ogawa, R. Pestotnik, L. Šantelj, T. Sumiyoshi, M. Tabata, M. Yonenaga, Y. Yusa: Construction of silica aerogel radiator system for Belle II RICH counter, Nucl. Instrum. Methods Phys. Res. A **876**, 129–132 (2017)

66. M. Adinolfi et al. (LHCb RICH Collaboration): Performance of the LHCb RICH detector at the LHC, Eur. Phys. J C **73**, 2431 (2013)

67. M. Aguilar-Benitez, L. Arruda, F. Barao, G. Barreira, A. Barrau, B. Baret, J. Berdugo, M. Buénerd, J. Casaus, I. Cernuda, D. Crespo, C. de la Guia, C. Delgado, C. Diaz, L. Derome, F. Giovacchini, P. Goncalves, A. Keating, G. Laurenti, A. Malinine, C. Mañá, J. Marín, G. Martinez, A. Menchaca-Rocha, R. Pereira, M. Pimenta, E. S. Seo, M. Vargas-Trevino: In-beam aerogel light yield characterization for the AMS RICH detector, Nucl. Instrum. Methods Phys. Res. A **614**, 237–249 (2010)

68. Y. K. Tanaka et al. ($\eta$-PRiME/Super-FRS Collaboration): Missing-mass spectroscopy of the $^{12}$C($p,d$) reaction near the $\eta$'-meson production threshold, Phys. Rev. C **97**, 015202 (2018)

69. J. J. Beatty, S. Coutu, M. Gebhard, N. Green, D. Hanna, B. Kunkler, M. Lang, I. Mognet, D. Müller, J. Musser, S. Nutter, N. Park, M. Schubnell, G. Tarlé, A. Tomasch, G. Visser, S. P. Wakely, I. Wisher: Cosmic-ray isotope measurements with HELIX, In: Proc. 35th Int'l Cosmic Ray Conf., (2017) PoS(ICRC2017)226